\documentstyle[12pt,psfig,obscite]{obsmag}

\normalbaselines
\title{\bf Probable detection of starlight reflected from the giant
	exoplanet orbiting {\boldmath $\tau$}~Bo\"{o}tis}

\author{
Andrew Collier Cameron$^*$, 
Keith Horne$^*$, 
Alan Penny$^\dagger$ and 
David James$^*$\\
$^*$ School of Physics and Astronomy, University of St Andrews,\\ North Haugh, 
St Andrews, Fife KY16 9SS\\
$^\dagger$ Rutherford Appleton Laboratory, Chilton, Didcot, Oxon OX11 0QX
}

\def\baselinestretch{1.0} 

\begin{document}

\maketitle

\begin{abstract}
Giant planets orbiting stars other than the Sun are clearly detectable 
through precise radial-velocity measurements of the orbital reflex 
motion of the parent star.  In the four years since the discovery\cite{mayor95_51peg} of the companion to the star 51 Peg, 
similar 
low-amplitude ``Doppler star wobbles'' have revealed the presence of 
some 20 planets orbiting nearby solar-type stars.  
Several of these newly-discovered 
planets\cite{butler97exoplanets,butler98,fischer99} 
are very close to their parent stars, in 
orbits with periods of only a few days.  Being an indirect technique, 
however, the reflex-velocity method has little to say about the sizes 
or compositions of the planets, and can only place lower limits on 
their masses.  Here we report the use of
high-resolution optical spectroscopy to achieve
a probable detection of the Doppler-shifted signature 
of starlight reflected from one of these objects, the giant 
exoplanet orbiting the star $\tau$ Bo\"otis.  
Our data give the planet's orbital inclination $i=29^\circ$,
indicating that its mass is some 8 times that of Jupiter, 
and suggest strongly that the planet has the size and 
reflectivity expected for a gas-giant planet.
\end{abstract}
 
A planet orbiting a star scatters back into space some of the 
starlight it receives.  To a distant observer, the ratio of the 
scattered flux $f_p$ from the planet to the direct flux $f_\star$ from 
its star depends on the planet's size and proximity to the star, which 
determine the amount of light intercepted, and on the scattering 
properties of the planet's atmosphere.  For a planet of radius $R_p$ 
in an orbit of radius $a$, the planet-to-star brightness ratio is 
$\epsilon \equiv f_p/f_\star = p(\lambda)(R_p/a)^2$, when
the planet is observed from above the sub-stellar point.
Here 
$p(\lambda)$ is the geometric albedo of the atmosphere for 
light of wavelength $\lambda$.   
The $\tau$~Boo planet\cite{butler97exoplanets} has a circular orbit
with a radius $a=0.0462 (M_{*}/1.2~M_{\odot})^{1/3}$~AU,
derived from Kepler's law with the 3.3~d orbital period 
and estimated stellar mass.  
If this is a giant planet, 
with a Jupiter-like size and albedo, we expect 
$\epsilon\sim10^{-4}$; the scattered starlight will be 10,000 to 
20,000 times fainter than the star even when viewed at the most 
favourable illumination angle $\alpha=0$.

Although $\tau$~Boo and its planet are never separated 
on the sky by more than 0.003 seconds of arc, this close-in
planet orbits its star with a relatively large velocity
$V_p = 152$~km~s$^{-1}$, based on observations 
of the star's reflex motion\cite{butler97exoplanets}.
The star and planet may therefore be easier to separate
by using the orbital Doppler shift than by direct imaging.
The pattern of photospheric absorption lines in the starlight
is preserved when the starlight is reflected from the planet,
apart from the Doppler shift due to the planet's orbital velocity,
and a multiplicative scaling by the wavelength-dependent albedo
of its atmosphere.
If the planet's orbit is inclined by an angle $i$ 
relative to our line of sight, its Doppler shift 
varies between $-K_p$ and $+K_p$ as it orbits around the star,
where $K_p=V_p\sin{i}$.  
For comparison, the star's absorption lines span $\pm 
28$~km~s$^{-1}$, largely due to the star's rotation.  Except in the 
unlikely case of a nearly face-on orbit, $i < 10^\circ$, the orbital 
Doppler shift should cleanly separate the planet's spectral lines from 
those of the star.


We observed $\tau$~Boo for a total of 48 hours on four nights in 1998 
April and five further nights in 1999 April, May and June using the 
Utrecht Echelle Spectrograph on the 4.2-m William Herschel Telescope 
at the Roque de los Muchachos Observatory on La Palma.  
The ranges in orbital phase covered by each night's observations are 
listed in Table \ref{tab:journal}.  


To isolate the feeble scattered-light spectrum of $\tau$~Boo's planet, 
we begin by constructing a high fidelity spectral model to accurately 
subtract the direct starlight from each of the 580 spectra.  
The initial stellar spectrum subtraction is a delicate operation.  
Small shifts of the spectrum on the detector and changes in the 
telescope focus and atmospheric seeing distort the spectral-line 
profiles from each exposure to the next.  Even small distortions of 
the strong stellar lines can produce changes larger than the faint 
signature of the planet.  We therefore calibrate these instrumental 
distortions by comparing each spectrum with a ``template'' spectrum 
$T(\lambda)$, which we construct by adding together all other spectra 
secured on the same night.  Making a new template for each night, 
rather than a single template for the whole observing run, helps to 
remove small changes in the detector response pattern from one night 
to the next.  The stability of the spectrograph is such that all 
spectra secured on a given night are mutually aligned to within 0.2 
pixel (0.6~km~s$^{-1}$), so that the template spectra are not 
appreciably broadened.  Our template spectra have a signal-to-noise 
ratio exceeding $10^4$ per spectral pixel in the best-exposed parts of 
the image.

We model each of the individually distorted stellar spectra as a 
linear combination of the template and its first and second 
derivatives with respect to wavelength 
$$ \displaystyle 
S(\lambda,\phi) = a_0 T(\lambda) + a_1 \frac{\partial 
T(\lambda)}{\partial \lambda} + a_2 \frac{\partial^2 
T(\lambda)}{\partial \lambda^2} \ .
$$
The line spread function (LSF) in our spectra is only 2.1 pixels wide 
(6.4~km/s), but the star's rotation spreads the stellar line profiles 
across 18 pixels (55~km~s$^{-1}$).  The above approximation is 
therefore adequate (in terms of stability against ``ringing'' and 
other related sampling effects) to correct the 0.2-pixel shifts and 
$\sim0.5$-pixel seeing-induced changes we see in width of the LSF. The 
coefficient $a_0$ serves to scale the template to match the observed 
spectrum, $a_1$ models the pixel shift between the spectrum and the 
template, and $a_2$ accounts for small changes in the width of the LSF 
due primarily to changes in atmospheric seeing.  We at first included 
higher derivatives to model changes in the skew and kurtosis of the 
LSF, but concluded that these effects were not significant.  To 
account for small changes in the LSF of the spectrograph across the 
focal plane, we represent the three coefficients as spline functions 
that vary smoothly across the frame.  We optimize the spline functions 
to fit each spectrum, using an iterative least-squares fit with 
outlier rejection, and then subtract the optimized model from each 
observed spectrum in turn.

This procedure very effectively removes the direct starlight signal,
leaving the planet signal deeply buried in noise. 
The procedure of constructing a template spectrum to model the direct 
starlight spectrum inevitably incorporates into the template some part 
of the planet signal.  This is particularly troublesome near the 
quadrature phases 0.25 and 0.75, when the planet's velocity is 
approximately constant.  Subtracting the template thus modifies and 
eliminates some of the planet signal. We account for this partial 
suppression later in the signal extraction process.

After subtracting the template to remove the stellar spectrum, we 
add together groups of four contiguous observations, primarily to 
save computer time in the subsequent analysis.  This reduces the 
number of spectra from 580 to 145, and boosts the signal-to-noise 
ratio to $\sim1200$ per spectral pixel.  Since the planet signal 
is expected to be at least $10^4$ times fainter than the starlight,
it remains deeply buried in noise at this stage of the analysis.


The planetary scattered-light signature is present in each spectrum as 
a faint Doppler-shifted copy of each of the star's spectral lines.  To 
build up this signal, we combine the velocity 
profiles of thousands of spectral lines using a 
least-squares deconvolution technique (LSD) that has become a standard 
tool for detecting and mapping stellar surface 
features\cite{donati97zdi}.

\begin{figure}[t]
\psfig{figure=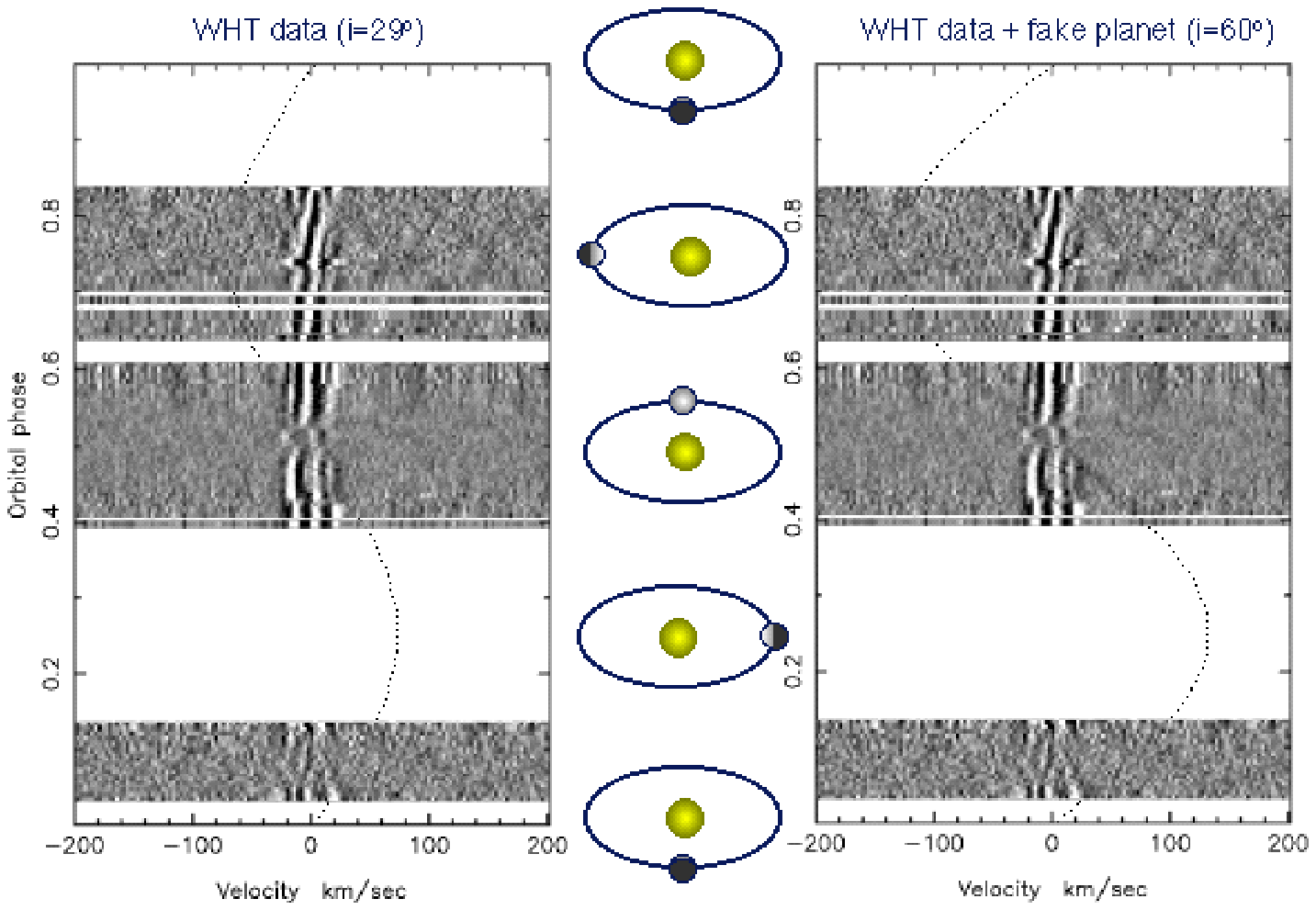,width=12.0cm} 

\caption[]{ Greyscale plots showing the 145 residual velocity profiles 
that arise from subtracting the stellar template model and averaging 
the profiles of $\sim2300$ spectral lines in each of the 145 spectra.  
The velocity scale is in the reference frame of the star, and time 
increases upward.  The orientation of the system at each orbital phase 
is sketched at centre.  The greyscale runs from black at $-10^{-4}$ to 
white at $+10^{-4}$ times the mean stellar continuum flux.  The dotted 
paths indicate the velocity curve of a planet orbiting with an orbital 
inclination of 29$^\circ$ (left-hand panel) and 60$^\circ$ (right-hand 
panel).  The right-hand panel shows the effect of adding the simulated 
spectrum of a planet with 1.4 Jupiter radii and a geometric albedo of 
0.55 to the original spectra.  The signature of this simulated planet 
appears as a dark linear feature crossing from $+80$~km~s$^{-1}$ at 
phase 0.4 to $-80$~km~s$^{-1}$ at phase 0.6.  The planetary signature 
detected in the data is much weaker, because of the low inclination, 
but would follow the velocity curve shown in the left-hand panel.  The 
``barber's pole'' pattern of travelling ripples lies wholly within the 
residual stellar profile and has a characteristic amplitude 2 to 
$4\times10^{-4}$ of the mean stellar continuum level.  Its form 
remains invariant in profiles deconvolved from independent subsets of 
the data at different wavelengths.  All data between phases 0.64 and 
0.83 were obtained in a 5-night interval during the 1998 season.  The 
ripple pattern appears both stronger and more coherent here than in
the 1999 dataset, with 5 nights spread over 2 months
selecting orbital phases near 0.5}
\label{fig:ph_decon60}
\end{figure}

Our LSD procedure uses a list of the known wavelengths and strengths 
of $\sim2300$ spectral lines that are present over the observed 
wavelength domain in a star of $\tau$ Boo's temperature and 
composition.  We determine, via $\chi^2$ minimization, a ``mean'' line 
profile which, when convolved with the known line pattern, yields an 
optimally-weighted match to the strengths and shapes of the lines in 
the observed spectrum.  This ``deconvolved profile'' is thus an 
average velocity profile that is representative of all the lines 
recorded in the spectrum, but with a vastly improved signal-to-noise 
ratio.  Blends are automatically compensated for, so that LSD gives a 
flat continuum outside the stellar profile, free of the sidelobes that 
are produced by simpler shift-and-add or cross-correlation procedures.  
We first apply LSD to the spectra of $\tau$~Boo, to establish the mean 
strength of its stellar absorption lines.  A second application of 
LSD, this time to the template-subtracted spectra, extracts the mean 
velocity profile of the planetary signal averaged over all of those 
lines.  The resulting template-subtracted velocity profiles, one for 
each spectrum, are displayed as a velocity-phase map in 
Fig.\ref{fig:ph_decon60}.

The most prominent pattern visible in Fig.~\ref{fig:ph_decon60} is a 
``barber's pole'' pattern at low velocities inside $\pm 
28$~km~s$^{-1}$.  This pattern is the same when we split the data into 
several wavelength regions, suggesting a stellar origin rather than a 
problem in the template registration and subtraction.  The ripples 
drift from the blue to the red wing of the profile in roughly half the 
orbital period of the planet.  One possible explanation would be the 
presence of solar-sized starspot groups crossing the face of the star as it 
rotates synchronously with the orbit.  The 
presence of large-scale inhomogeneities in the photospheric velocity 
field\cite{toner88} could also give low-amplitude, 
rotationally-modulated distortion of the line profiles without 
producing significant optical variability.  A rotational modulation of 
the star's chromospheric Ca~II H \&\ K emission has been detected with 
a period of 3.3~days\cite{baliunas97}, essentially the same as the 
3.312567-day orbital period of the planet.  The barber's-pole pattern 
thus provides useful independent confirmation that the star does 
indeed rotate synchronously with the planet's orbit.

Outside the range $\pm 28$~km~s$^{-1}$, the residuals have the general 
appearance of random noise, but with some correlation in time due to 
nights with residual flat-fielding errors.  The root-mean-square 
scatter indicates a signal-to-noise ratio of $\sim3.5\times10^4$ per 
spectral pixel relative to the original stellar continuum level.  This 
noise level is 10\% to 15\% greater than photon noise estimates
propagated through the data extraction and deconvolution.
LSD has therefore 
realized a 30-fold increase in signal-to-noise ratio by combining the 
velocity profiles of the $\sim2300$ spectral lines.  This should be 
just sufficient to detect the brightest expected planet signature in 
those of the 145 LSD velocity profiles taken at gibbous phases when 
the planet is Doppler-shifted well clear of the stellar profile.

To search for candidate planet signals, measure their strengths and 
assess their significance, and to quantify upper limits on 
planets at other inclinations, we must extract information from the 
LSD velocity profiles at all 145 orbital phases.  We do this in an 
optimal way by using a matched filter to account for the expected 
changes in the planet's Doppler shift and brightness with orbital 
phase.

We model the velocity profile of a planet signal as a moving Gaussian
$$
G(v,\phi,K_p) = \frac{ W g(\phi,i) }{ \sqrt{\pi} \Delta }
{\rm exp} \left\{ \displaystyle
- \left( \displaystyle
\frac{ v - K_p \sin{\phi} }{\Delta} \right)^2
\right\} 
\ .
$$
The velocity width parameter $\Delta$ is chosen to 
match the expected averaged width of the stellar lines reflected from 
the planet. 
Because the star rotates synchronously with the planet's orbit, the 
planet sees the stellar spectrum without the rotational broadening
that we see in the direct starlight\cite{charbonneau98}. 
We estimate the width parameters of the direct starlight and the reflected
light to be 13.2 and 6.4~km~s$^{-1}$ respectively, from Gaussian fits to the deconvolved profiles of $\tau$~Boo and 
the slowly rotating giant star HR~5694. The latter  
has an F7~III spectral type (temperature) and elemental abundances 
similar to $\tau$~Boo\cite{baliunas97}.
The larger width and 
shallower depth of $\tau$~Boo's lines, due largely to the star's axial 
rotation, yields\cite{gray82tboovsi,fuhrmann98,gonzalez98} a projected 
equatorial rotation speed $v\sin i=14.8\pm 0.3$~km~s$^{-1}$, where $i$ 
is the inclination of the spin axis to the line of sight.

The line strength parameter $W$ is set to 
match that of $\tau$~Boo.  For $g(\phi,i)$, which modulates the 
planet's brightness with orbital phase, we adopt an empirically-determined 
polynomial approximation to the phase function of Venus\cite{hilton92}.  
The orbital velocity 
amplitude is $K_{p} = V_p \sin{i}$, where $V_p = 152$~km~s$^{-1}$.  
Note that $K_p$ and $i$ are not independent parameters; we compute $i$ 
for each value of $K_p$.

To compensate for the attenuating effect of a blurred planet signal 
being present in the template, we mimic the effect in constructing
the matched filter $M(v,\phi,K_p)$. We do this by subtracting the 
flux-weighted average of the Gaussian planet signals on each night, 
emulating precisely the attenuation arising from the template 
construction and subtraction.

To allow for the possibility that the planet's albedo is wavelength 
dependent, we subdivide our data into 6 independent wavelength ranges 
and measure the planet signal strength $\epsilon(\lambda)$ in each 
data subset.  Thus we construct template-subtracted LSD velocity 
profiles $f(v,\phi,\lambda)$, and their corresponding noise variances 
$\sigma^2(v,\phi,\lambda)$, for 6 independent subsets of the echelle 
orders at wavelengths $\lambda$.  For any set of trial values of the 
parameters $K_p$ and $\epsilon(\lambda)$, we quantify the ``badness of 
fit'' to the data by means of the standard $\chi^2$ statistic
$$
\chi^{2} \equiv
\sum_v \sum_\phi \sum_\lambda 
\left(
\frac{ f(v,\phi,\lambda) - \epsilon(\lambda) M(v,\phi,K_p) }
{ \sigma(v,\phi,\lambda) }
\right)^2
\ .
$$
Thus we scale the matched filter $M(v,\phi,K_p)$ by factors 
$\epsilon(\lambda)$ to fit LSD velocity profiles 
$f(v,\phi,\lambda)\pm\sigma(v,\phi,\lambda)$, which are like those 
shown in Fig.~\ref{fig:ph_decon60} but constructed from 
wavelength-restricted subsets of the data.  The best fit minimizes 
$\chi^{2}$, yielding the optimal estimates of $K_p$ and 
$\epsilon(\lambda)$.  The increase in $\chi^{2}$ for nearby parameter 
values, $\Delta \chi^2$, is used to judge different models, their 
relative probabilities being proportional to ${\rm exp}\left( -\Delta 
\chi^2/2 \right)$.  To prevent spurious planet signatures arising from 
the ``barber's pole'' pattern in the residual stellar profile,
we exclude from the fitting procedure
all pixels within 31.5~km~s$^{-1}$ of the centre of the 
stellar profile.

We verified that our procedure is capable of revealing faint planetary 
signals in the presence of realistic noise levels and patterns by 
adding a simulated planetary signal to the observed spectra, then 
repeating the template construction, subtraction and LSD analysis. 
The planet was simulated by the spectrum of HR 5694, Doppler 
shifted to the appropriate orbital velocities and scaled according 
to the phase function expected for a planet with a radius 1.4 times 
that of Jupiter and a  grey albedo $p=0.55$ (Fig.~\ref{fig:ph_decon60}).

In searching for candidate planetary signals in the WHT data, we begin by testing the null hypothesis that no planet is present, in 
comparison with the alternative hypothesis that a ``grey'' planet with 
wavelength-independent albedo is present at some value of $K_p$.  We 
probe for evidence of a planet over the range 40~km~s$^{-1} < K_p < 
152$~km~s$^{-1}$, corresponding to inclinations $15^\circ < i < 
90^\circ$.  At lower velocities and inclinations the planet's orbit 
velocity never emerges from the low-velocity region affected by the 
``barber pole'' pattern. 
An inclination than $i=15^\circ$ would also
require an implausibly high stellar equatorial rotation 
speed\cite{charbonneau99tauboo}.

\begin{figure}[t]
\psfig{figure=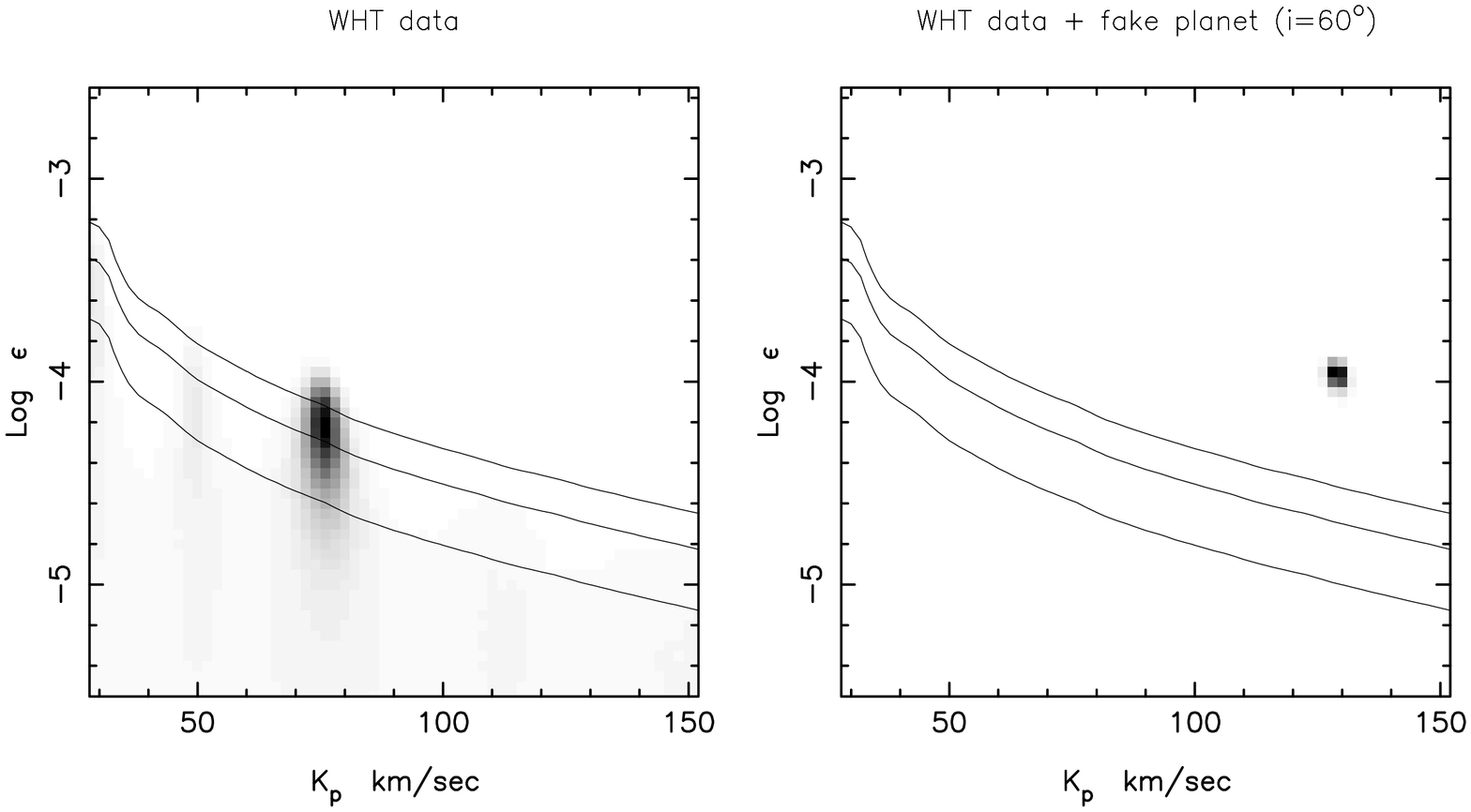,bbllx=100pt,bblly=71pt,bburx=476pt,bbury=755pt,angle=-90,width=14cm} 
\caption[]{The improvement in the fit to the data is shown for the 
assumption of a planet with a wavelength-independent (grey) albedo.  
The two panels present identical analyses of the residual velocity 
profiles shown in the corresponding panels of 
Fig.~\ref{fig:ph_decon60}, representing the WHT data alone (left) and 
the WHT data after adding a simulated planet signal for $i=60^\circ$ 
(right).  For each possible value of the unknown planet-to-star 
brightness ratio $\epsilon$, and of the planet's projected orbit 
velocity $K_p$, darker shades denote progressively better fits to the 
data.  The probability relative to the best-fit model increases from 0 
for white to 1 for black.  In the right panel, the fake planet's 
parameters $K_p=132$~km~s$^{-1}$ and $\epsilon=1.07\times10^{-4}$ are 
correctly recovered.  In the left panel, the best fit to the WHT data 
occurs for $K_{p}=74$~km~s$^{-1}$ and $\epsilon=7.5\times10^{-5}$.  
The ``grey-planet'' model significantly improves the fit (darker 
shading) relative to the ``no-planet'' model ($\epsilon=0$); the 
improvement in $\chi^{2}$ after optimizing $\epsilon$ and $K_{p}$ is 
$\Delta\chi^{2}=9.74$.  If no planet were present, the probability of 
such an improvement occurring by chance should have a $\chi^{2}$ 
distribution with 2 degrees of freedom.  The actual distribution, 
derived for the noise and systematic error in our data via a bootstrap 
error analysis, has a slightly longer tail, so that 
$\Delta\chi^{2}=9.74$ rejects the ``no-planet'' hypothesis with 97.8\% 
confidence.  If no planet is present, the best-fit model should lie 
below the 68.3\%, 95.4\%, and 99.7\%\ bootstrap confidence limits 
shown as solid curves.  These limits decrease with $K_p$ because for 
larger $K_p$ the planet lines move more rapidly across the stellar 
profile and into the region covered by the data between phases 0.4 and 
0.6.  The light grey shading below the confidence limits is consistent 
with noise, and planets are strongly excluded in the white zones above 
the highest confidence limit.
}
\label{fig:limits_noplanet}
\end{figure}

Fig.~\ref{fig:limits_noplanet} presents the relative probability map 
as a function of $K_p$ and $\epsilon$, showing significant evidence 
for a planet at $K_{p}=74\pm3$~km~s$^{-1}$ and $\epsilon=7.5\pm3\times 
10^{-5}$.  The ``grey-planet'' hypothesis introduces 2 parameters, 
$\epsilon$ and $K_p$.  The improvement in the fit, reducing $\chi^2$ 
by $\Delta\chi^2=9.74$ with only 2 degrees of freedom, rejects the 
``no-planet'' hypothesis with 97.8\% confidence, based on a bootstrap 
analysis of the error distribution.  Thus if no planet is present, the 
probability is about 2\%\ that the noise in our data would produce a 
spurious signal this strong. When the synthetic planet 
signature at $i=60^\circ$ is injected into the data, its velocity 
amplitude and strength are recovered correctly 
(Fig.\ref{fig:limits_noplanet}).

\begin{figure}[t]
\psfig{figure=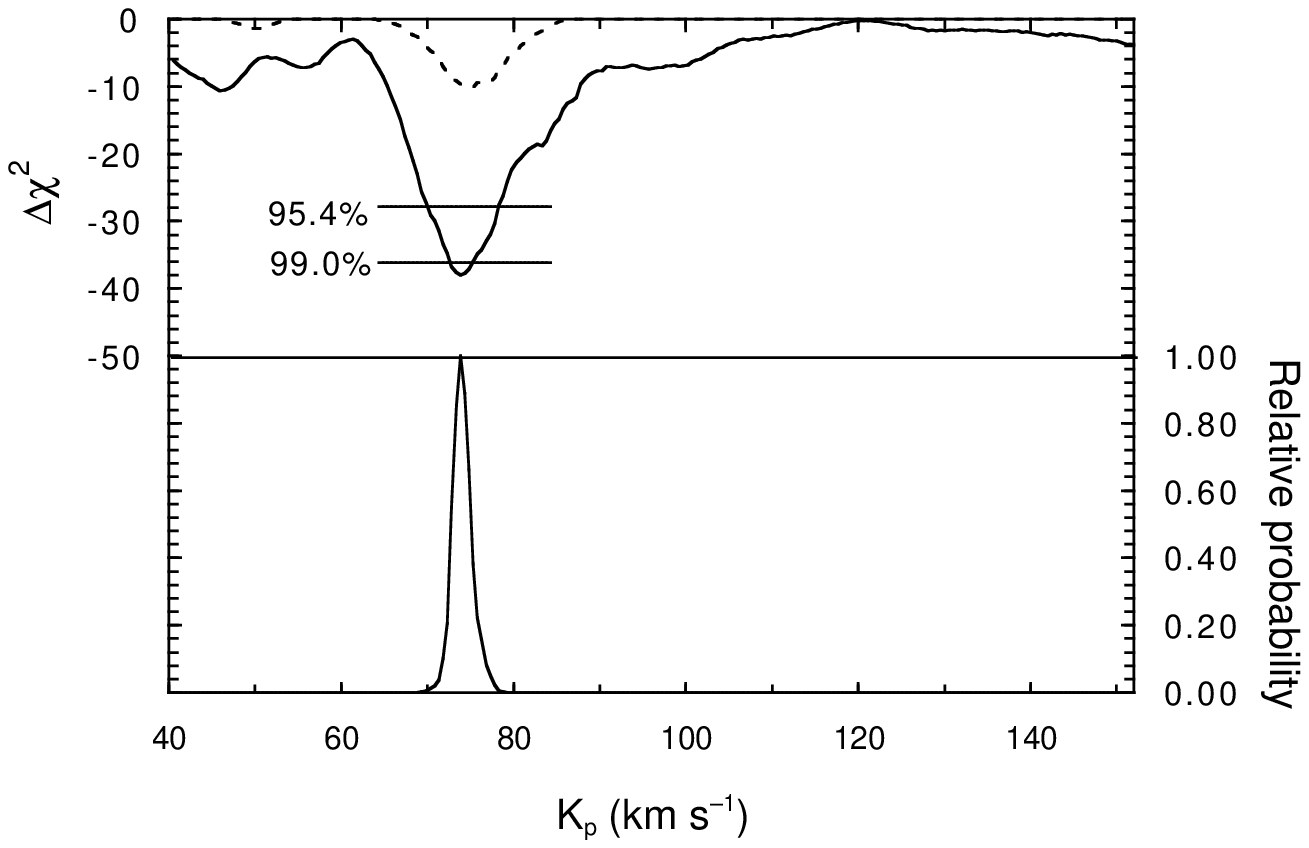,bbllx=0pt,bblly=0pt,bburx=350pt,bbury=230pt,width=8.6cm} 
\caption[]{ The evidence for a planet in the WHT data is assessed as a 
function of the planet's orbit velocity $K_p$.  The top panel gives 
$\Delta \chi^2$, the reduction in the $\chi^2$ of the fit relative to 
the assumption that no planet is present, for the 2-parameter model in 
which the planet is assumed to have constant (grey) albedo (dashed 
curve), and for the 7-parameter model with the planet's albedo varying 
with wavelength (solid curve).
The horizontal lines give for two 
values of $\Delta \chi^2$ the corresponding probabilities that the 
improvement in $\chi^2$ is too large to be attributed to noise.  
The 2-parameter constant albedo model
gives a significant (97.8\%) improvement at $K_p=74$~km~s$^{-1}$,
and the wavelength-varying albedo model with 7 fitted parameters
raises this probability to 99.2\%.  
The lower panel shows the probability of 
the 7-parameter model as a function of $K_p$, scaled to 1 for the 
best-fit model $K_p=74\pm2$~km~s$^{-1}$.  }
\label{fig:mchsq_all}
\end{figure}

We examine next the hypothesis that the albedo depends on wavelength, 
by testing the ``no-planet'' hypothesis against a model with 7 
additional parameters, $K_p$, and $\epsilon(\lambda)$ for 6 
wavelengths $\lambda$.  Thus we split the data into six independent 
wavelength bands chosen so that the recorded flux is divided 
roughly equally 
among the six bands.  By fitting independent values of $\epsilon$ to 
different wavelength ranges we allow for the possibility that the 
planet's albedo spectrum could contain broad absorption bands due to molecular 
species such as TiO overlying the main scattering layers.  
The greatest improvement in 
$\chi^{2}$ occurs again at $K_{p}=74$~km~s$^{-1}$ 
(Fig.~\ref{fig:mchsq_all}).  With 
$\Delta\chi^{2}=37.96$, the ``no-planet'' hypothesis is again 
rejected, this time with 99.2\%\ confidence for 7 degrees of freedom, 
an 0.8\%\ probability of occurring by chance.

This result implies a significant departure from a 
wavelength-independent albedo spectrum. 
The signal is concentrated in the three wavelength bands from 456~nm 
to 524~nm,
with the central 479 to 509~nm band contributing half of the signal.
Averaging over these three bands alone, we obtain 
$\epsilon = 1.9\pm0.4\times10^{-4}$.
No significant signal is present in the other bands redward and
blueward of these three.  

We conclude that there is strong evidence in our data for a planetary 
signal at $K_{p}=74$~km~s$^{-1}$, which cannot easily be explained as a 
spurious detection caused by the photon statistics and locally 
correlated errors in our data.  


The results obtained so far have relied on calculations of the 
planet's orbital phase based on
an orbit period $P=3.312567$~d and conjunction epoch $T_0$
kindly communicated to us by G.~Marcy.
These parameters were derived from high-precision studies
of the star's velocity, and we are therefore justified in
holding $P$ and $T_0$ fixed at their known values
while searching for reflected-light signals
over the feasible range of the unknown orbit velocities $K_{p}$.  
If the planet signal is strong enough, however, we should be 
able to measure $P$ and $T_0$ independently from our data,
and those values should be consistent with the known values.
Considering a small range around Marcy's values, 
our data yield $P=3.3128\pm0.0002$ days ($1\sigma$ errors),
consistent with Marcy's more accurate period,
and a conjunction phase $\phi_0 = 0.007\pm0.003$,
consistent with Marcy's $0.000\pm0.002$.
Our measurement of $K_p=74\pm2$~km~s$^{-1}$ is also insensitive
to small changes in the period and epoch of the planet's orbit.  

We also considered different values for the velocity width 
$\Delta$ of the absorption lines in the light scattered from the 
planet.  The best fit was found for $\Delta = 6\pm2$~km~s$^{-1}$.
This is in good agreement with the 6.4~km~s$^{-1}$ width of the lines in the 
spectrum of HR~5694, confirming that the planet sees the starlight
without rotational broadening, and hence that the star rotates 
synchronously with the planet.
   
\begin{figure}[t]
\psfig{figure=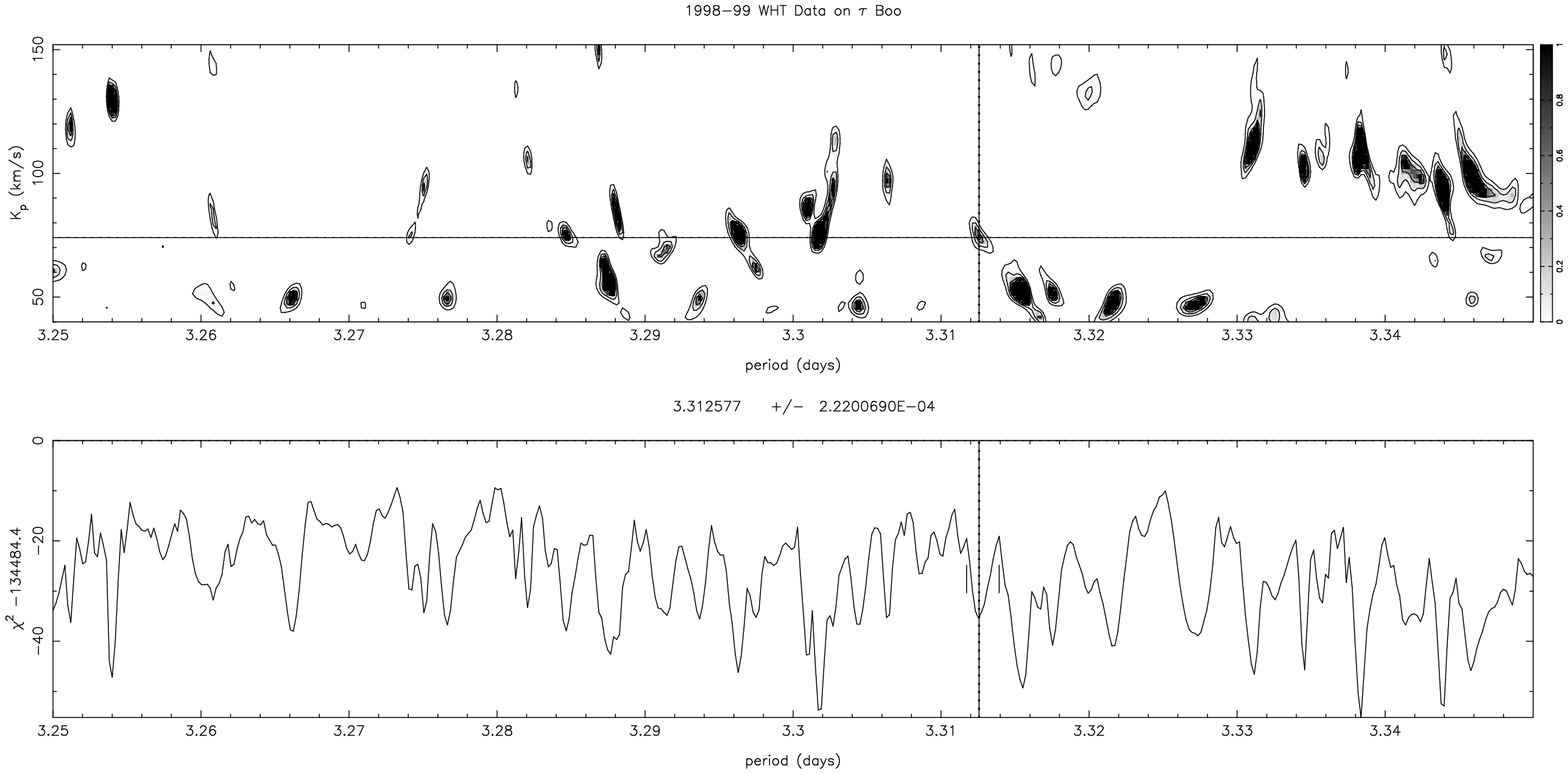,angle=-90,width=16cm}
\caption[]{  This periodogram analysis searches for
evidence of reflected light from the $\tau$~Boo planet
over a range of trial values of the orbit velocity $K_p$
and orbit period $P$.
For each value of $K_p$ and $P$, the 6 values $\epsilon(\lambda)$
are optimized to fit 6 independent wavelength subsets
of the echelle data.
The lower panel shows the best $\chi^2$ found for $K_p$ between 40 and
152~km~s$^{-1}$ at each trial period $P$.
In the upper panel, the greyscale indicates for each value of $K_p$ and $P$
the relative probability of the best-fit model,
with white to black representing increasing relative probability.
The vertical line marks the accurate value of $P$,
from G. Marcy's analysis of the star's Doppler signal.
Our detection is the probability peak occurring 
at the correct period near $K_p=74$~km~s$^{-1}$,
as marked by the intersection between the vertical and
horizontal lines.
The contours give 1$\sigma$, 2$\sigma$, and 3$\sigma$
confidence regions for this peak, 
based on $\Delta\chi^2$ for 2 degrees of freedom.
A number of other peaks, some giving even better fits to the 
data than our detection,
are seen at different values of $P$ and $K_p$.
These peaks arise mainly from noise in the data. We infer from 
the number of such peaks that there is a
3 to 5\%\ probability that our detection is a spurious noise peak
that happens by chance to coincide with the known period.
Note that some periodogram features appear twice,
shifted horizontally by 0.03 days. These
correspond to pairs of periods that differ by 1 orbital cycle per year,
due to the 1 year gap that separates our 1998 and 1999 observations.
}
\label{fig:periodogram}
\end{figure}

As an additional test, we use a periodogram analysis to estimate
the probability that our planet signal could be an artifact of
noise features that happen by chance to line up along the
planet's sinusoidal path in the velocity-phase diagram.  
For this test, we relax our knowledge of the planet's orbital period $P$,
and re-analyze the data for many different trial periods
in the range $3.25 < P < 3.35$ days.
We estimate the ``false-alarm'' probability as the fraction of
periods that yield a planet signal stronger than our candidate detection
for some value of $K_{p}$ in the feasible range.
In order to keep the orbital phases close to 0.5 in the 1999 data
for all periods considered, we set the time of zero phase to an epoch
of conjunction just prior to 1999 May 5.
The pattern of spurious peaks as a function of 
both $K_p$ and $P$ is then as shown in Fig.~\ref{fig:periodogram}.  
The lower panel of this figure shows the minimum value 
of $\chi^{2}$, optimised over the range $40<K_p<152$~km~s$^{-1}$, 
at each period.  Of all the 
periods sampled, a small fraction yield spurious detections stronger 
than our planet signal, and close enough to their local 
maxima to pass the same local tests as our candidate signal.  

We conclude from the periodogram test that the probability of 
encountering a spurious peak
with $K_{p}> 74$ km s$^{-1}$ is between 3 and 5 percent.
This is our most pessimistic assessment of the 
probability that our detection is spurious.  
If other prior knowledge is taken into account -- in particular, 
the restrictions imposed on the orbital inclination by
the star's synchronous rotation --
the false-alarm probability is reduced further still.
 

The best-fitting orbital velocity amplitude $K_p=74$~km~s$^{-1}$ yields
an orbital inclination $i=29^\circ$ and implies a planet mass 
eight times that of Jupiter, twice the minimum value for an edge-on 
orbit.  The low inclination is consistent with arguments for 
synchronous rotation, based on the star's equatorial rotation speed, 
rotation period, angular diameter and 
distance\cite{baliunas97,fuhrmann98}, plus the requirement that the 
star's rotational synchronisation timescale 
($1.2\times10^{11}\sin^{8}i$ years) must then be shorter than its 
present age (less than $\sim3$ Gyr)\cite{marcy9751peg,drake98sync}.

At this low inclination, the fact that we are able to detect the 
planet at all suggests a large radius and/or reflectivity.
Indeed, our planet-to-star flux ratio 
$\epsilon = 1.9\pm0.4\times10^{-4}$, 
averaged between 456 and 524~nm, 
appears to conflict with recent work\cite{charbonneau99tauboo}
by Charbonneau et al., who establish an upper limit
on the brightness of the $\tau$~Boo planet 
$\epsilon<10^{-4}$ with 99.9\%\ confidence 
for $i\sim30^\circ$ at similar wavelengths.
We caution, however, that 
the value of $\epsilon$ produced by our fitting method is sensitive to 
the Venus-like form we adopt for the phase function.  A lower value 
of $\epsilon$ and hence a smaller radius results if the true phase 
function is less strongly peaked at zero phase angle than the one used 
here.
We also note that, while Charbonneau et al.\ 
estimate a $\sim20$\% attenuation of planet signals
due to their template-subtraction process,
their tests recovering simulated planet signatures 
were incomplete because the template spectrum was fitted 
before rather than after adding the planet.  
We designed our 
matched-filter fitting function to mimic closely the effects of template 
attenuation. Our tests with recovery of synthetic planet 
signatures indicate that this was adequate to compensate for the 
effect.  We conclude that differences in the treatment of the phase 
function and template attenuation effects are sufficient to reconcile 
the apparent conflict between our detection and the upper limit 
published by Charbonneau et al.

The radius we infer for $\tau$~Boo's planet is 1.8 times that of Jupiter,
assuming a Jupiter-like albedo $p=0.55$,
and using the Venus-like phase function.
This is somewhat larger than the radii predicted by recent structural 
and evolutionary models\cite{guillot97abos}, which range 
from $\sim1.4$~R$_{J}$ at age 2~Gyr to  $\sim1.1$~R$_{J}$ at 3~Gyr.
If we use a Lambert-sphere model, as employed by Charbonneau et al., 
our measured radius decreases by 12\% to $1.6~R_J$.

Our candidate detection of starlight scattered from the atmosphere 
of an extra-solar planet strengthens the case for the existence of the giant, 
close-orbiting planets whose presence has so far been inferred only 
indirectly from the reflex motions of their parent stars.  The 
strength of the detection indicates that the $\tau$~Boo planet
is a gas giant, and suggests that its optical reflectivity may be 
appreciable only over a narrow range of blue-green wavelengths.  Further 
observations will be needed to determine the form of the phase 
function, and so improve estimates of the radius and albedo.  We 
envisage that it will be possible to infer the presence or absence of 
molecular species such as TiO, methane and water in the vapour phase, 
by carrying out more sensitive studies over suitably defined 
wavelength ranges.  Finally, we note that the close-orbiting planets 
of other systems, including both 51~Pegasi and the $\upsilon$~Andromedae triple-planet system,
should be amenable to similar studies in the near future.

{\bf Acknowledgments:} This work is based on observations secured with 
the 4.2-m {\em William Herschel Telescope} at the Observatorio del 
Roque de los Muchachos on La Palma, and used {\em Starlink}-supported 
hardware and software.
We thank the referees for helpful criticism.

Correspondence should be addressed to A.C.C. (e-mail 
andrew.cameron@st-and.ac.uk).

\newpage

\begin{table}
\caption{Journal of observations.  The UTC mid-times and orbital 
phases are shown for the first and last groups of four spectra secured 
on each night of observation. The spectra spanned the optical wavelength 
range from 385~nm to 611~nm at a resolving power $\lambda/\Delta\lambda=49,000$.  The signal-to-noise ratio, typically $\sim600$ in each 3~km~s$^{-1}$ 
wide spectral pixel, was limited by photon statistics. The number of 
groups of four such consecutive spectra is given in the 
final column.  The orbital phase $\phi$ increases from 0 to 1 around 
the orbit, and is defined so that the planet is closest to the 
observer at phase 0.  In the 1999 season we concentrated 
on the phase range between 0.4 and 0.6, where the planet is expected 
to be brightest. The orbital phases are computed from the epoch 
of inferior conjunction $T_0$ at HJD~2451269.756 and orbital period 
$P=3.312567$~d, kindly communicated to us directly by G. Marcy.  
}
\label{tab:journal}
\begin{tabular}{cccccc}
\hline
UTC start             & Phase   &  UTC End             & Phase  & 
Number\\
                      &         &                      &        & 
of groups\\
                      &         &                      &        &\\
\hline \hline
1998 Apr 09  22:09:43 &	0.4254	& 1998 Apr 10 05:37:05 & 0.5192	&25\\
1998 Apr 10  22:04:40 & 0.7262  & 1998 Apr 11 06:20:54 & 0.8303 &29\\
1998 Apr 11  22:09:28 & 0.0291  & 1998 Apr 12 05:58:44 & 0.1275 &21\\
1998 Apr 13  23:02:54 & 0.6441  & 1998 Apr 14 06:08:23 & 0.7332 &9\\
                      &         &                      &        &\\
1999 Apr 02  22:06:45 & 0.4931  & 1999 Apr 03 06:08:47 & 0.5941 &15\\
1999 Apr 25  21:43:17 & 0.4411  & 1999 Apr 26 05:31:59 & 0.5393 &10\\
1999 May 05  21:56:59 & 0.4628  & 1999 May 06 04:51:57 & 0.5497 &15\\
1999 May 25  20:59:45 & 0.4884  & 1999 May 26 03:56:18 & 0.5757 &11\\
1999 May 28  21:06:19 & 0.3954  & 1999 May 29 02:40:44 & 0.4655 &10\\
\hline
\end{tabular}
\end{table}

\def\baselinestretch{1.0} 
\newpage

\newpage

\newpage

\newpage

\end{document}